\documentclass[%
superscriptaddress,
 reprint,
 amsmath,amssymb,
 aps,
onecolumn]{revtex4-2}

\usepackage{graphicx}
\usepackage{dcolumn}
\usepackage{bm}
\usepackage{comment}
\usepackage{pgfplots}

\definecolor{amaranth}{rgb}{0.9, 0.17, 0.31}
\begin{document}

\preprint{APS/123-QED}

\title{\LARGE Conservation of a Half-Integer Angular Momentum in Nonlinear Optics with a Polarization Möbius Strip}

\author{Martin Luttmann}\email{martin.luttmann@cea.fr}
\affiliation{%
Universit\'e Paris-Saclay, CEA, CNRS, LIDYL, 91191 Gif-sur-Yvette, France
}

\author{Mekha Vimal}%
\affiliation{%
Universit\'e Paris-Saclay, CEA, CNRS, LIDYL, 91191 Gif-sur-Yvette, France
}

\author{Matthieu Guer}
\affiliation{%
Universit\'e Paris-Saclay, CEA, CNRS, LIDYL, 91191 Gif-sur-Yvette, France
}
\affiliation{Grupo de Investigación en Aplicaciones del Láser y Fotónica, Departamento de Física Aplicada, University of Salamanca, Salamanca E-37008, Spain}

\author{Jean-François Hergott}
\affiliation{%
Universit\'e Paris-Saclay, CEA, CNRS, LIDYL, 91191 Gif-sur-Yvette, France
}

\author{Antonio Z. Khoury}
\affiliation{Instituto de Física, Universidade Federal Fluminense, 24210-346 Niterói, RJ, Brazil}


\author{Carlos Hern\'andez-Garc\'ia}
\affiliation{Grupo de Investigación en Aplicaciones del Láser y Fotónica, Departamento de Física Aplicada, University of Salamanca, Salamanca E-37008, Spain}

\author{Emilio Pisanty}
\affiliation{Department of Physics, King's College London, Strand Campus, WC2R 2LS, London, UK}

\author{Thierry Ruchon}
\email{thierry.ruchon@cea.fr}
\affiliation{%
Universit\'e Paris-Saclay, CEA, CNRS, LIDYL, 91191 Gif-sur-Yvette, France
}

\date{\today}

\begin{abstract}
\textbf{Symmetries and conservation laws of energy, linear momentum and angular momentum play a central role in physics, in particular in nonlinear optics. 
Recently, light fields with non trivial topology, such as polarization Möbius strips or torus-knot beams, have been unveiled. They cannot be associated to well-defined values of orbital and spin angular momenta (OAM and SAM), but are invariant under coordinated rotations, i.e. rotational symmetries that are generated by the generalized angular momentum (GAM) operator, a mixture of the OAM and SAM operators. The discovery of the GAM, which at variance with integer-valued OAM and SAM, can have arbitrary value, 
raises the question of its conservation in nonlinear optical processes.   
By driving high harmonic generation with a polarization Möbius strip and implementing novel OAM characterization methods in the XUV range, we experimentally observe the conservation of the GAM, each harmonic carrying a precise half-integer GAM charge equal to that of the fundamental field multiplied by the harmonic order. The GAM is thus revealed as the appropriate quantum number to describe nonlinear processes driven by light fields containing topological polarization singularities.} 
\end{abstract}
\maketitle

Beyond bosons and fermions, which respectively carry integer and half-integer values of spin angular momentum, quantum particles whose motion is confined in two dimensions can exhibit intruiguing angular momentum quantizations. For instance, electrons can obey "anyonic" (i.e., neither bosonic nor fermionic) statistics in the fractional quantum Hall effect \cite{Arovas1984}. Similarly, as one particular direction is determined by their propagation axis, paraxial light beams do not show any full 3D rotational symmetry. Ballantine \textit{et al.} \cite{Ballantine2016} recently demonstrated that even though photons are classified as bosons, light beams that are superposition states of different OAM and SAM can be described in terms of an angular momentum operator with non-integer eigenvalues. For instance, monochromatic light fields that are invariant under a rotation of the spatial dependence by an angle $\varphi$, followed by a rotation of the polarization vector by a fraction $\gamma \varphi$ of that angle - a transformation called coordinated rotation \cite{Pisanty2019Nature} - are eigen-vectors of the generalized angular momentum (GAM): $J_{z, \gamma} = L_z+\gamma S_z$, with $J_{z, \gamma},$ $L_z$ and $S_z$ the projections of the GAM, OAM and SAM operators along the propagation axis $z$. Depending on the value of $\gamma$, $J_{z, \gamma}$ has integer or half-integer eigen values, the latter yielding a fermionic-like spectrum. Coordinated-rotations-invariant monochromatic light fields are topologically equivalent to twisted ribbons or Möbius strips  \cite{Freund2005, Freund2009, Galvez2017}. In the more general case of polychromatic fields,  the eigen-values of $J_{z, \gamma}$ can be arbitrary numbers. The topology of the corresponding light beam and of the symmetry group is that of a torus knot \cite{Pisanty2019Nature}, so that the GAM can be identified as a torus-knot angular momentum. This formalism provides a relevant alternative to the usual descriptions relying on "fractional OAM" states \cite{Gotte2008, Turpin2017}. Note that in the specific case of single-color beams, the GAM can also be expressed in terms of the Pancharatnam topological charge \cite{delasHeras2022}. The conservative nature of the GAM is expected to find a rich scenario in nonlinear optics \cite{Pisanty2019, Turpin2017}. However the experimental verification of the GAM conservation raises specific difficulties and has not been reported yet.

\begin{figure*}
    \centering
    \includegraphics[width=17cm]{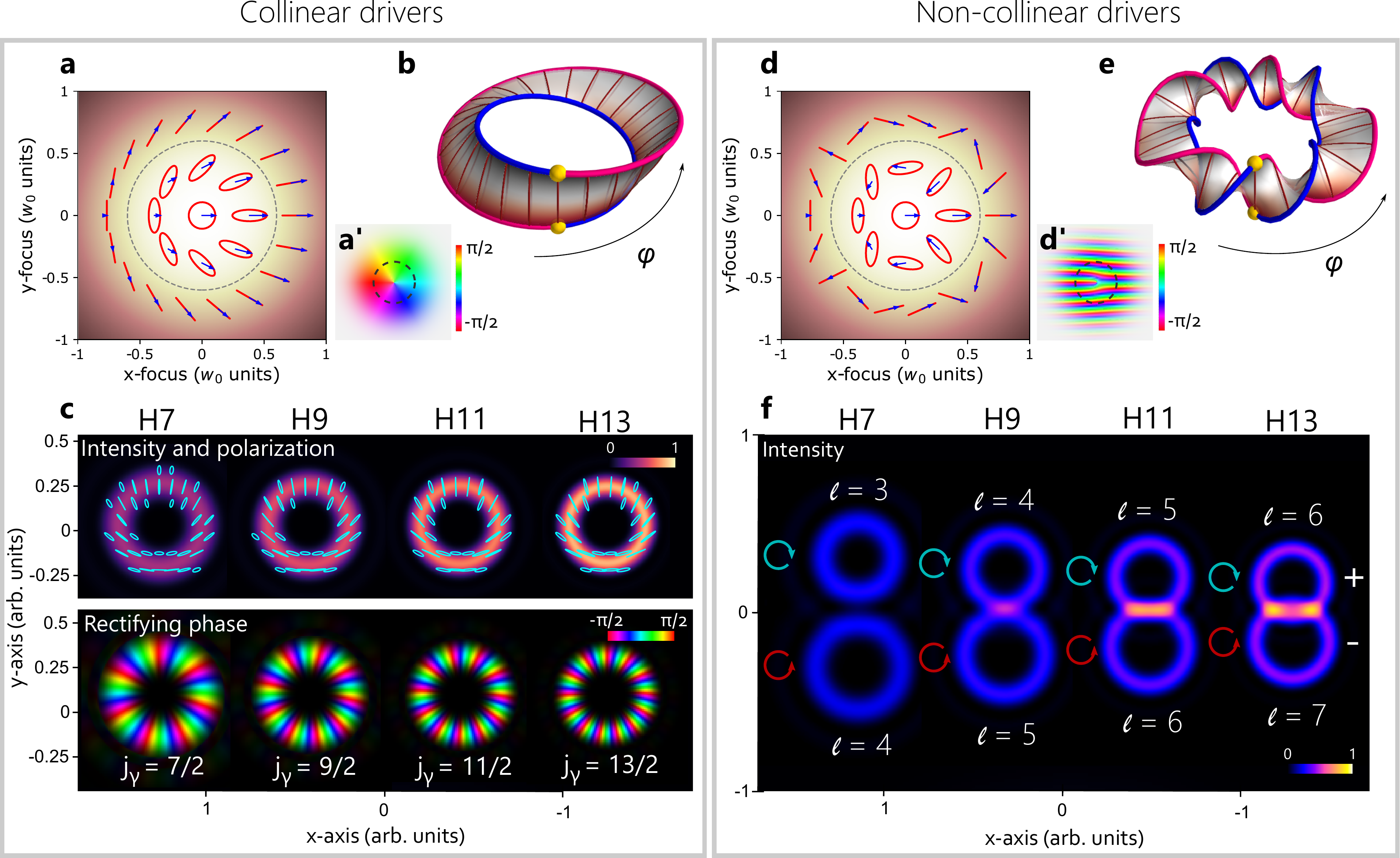}
    \caption{\textbf{Möbius strip topology of the driving IR field and corresponding XUV spectra, for collinear (left column) and non-collinear (right column, $\theta = 50$ mrad) drivers.} \textbf{(a, d)} Polarization state of the total IR driving field (red curves) and electric field at time $t=0$ (blue arrows), for three different raddi. 
    The background colormap displays the field intensity. \textbf{(a', d')} Orientation angle of the IR polarization ellipse.  \textbf{(b, e)} Möbius strip constructed by stacking the IR polarization ellipses along the dashed circle in A,D and connecting those corresponding to $\varphi =0$ and $\varphi=2\pi$. The blue and pink lines trace the path of the tip of the ellipse over a $2\pi$ range of the azimuthal angle $\varphi$. The yellow spheres indicate the connecting points of these two curves. \textbf{(c)} Simulated far field harmonics. The top panel shows the intensity and polarization state (cyan ellipses). The bottom panel displays the rectifying phase, defined as $\frac{1}{2}\text{arg} (\textbf{E}\cdot \textbf{E})$, where $\textbf{E}$ is the complex Fourier amplitude of the harmonic. The GAM charge $j_\gamma$ corresponds to the number of azimuthal $\pi$ phase jumps divided by 2. \textbf{(f)} Simulated far field XUV intensity. The blue (resp. red) arrows represent $\sigma = +1$ (resp. $\sigma = -1$) SAM states. }
    \label{fig:fig1}
\end{figure*}

In this Article, we demonstrate the conservation of the GAM in a nonlinear optical process. Such investigation would be difficult to carry out with nonlinear crystals, which are generally birefringent. The rotational invariance would thus be broken, leading to a transfer of angular momentum between light and matter, or optical spin-orbit coupling \cite{Bliokh2015, DaSilva2022}. Alternatively, if the conversion process is mediated by an isotropic medium, optical OAM and SAM are conserved independently. This is precisely achievable in high-order harmonic generation (HHG), a highly non-perturbative nonlinear optical process occuring in gases \cite{Villeneuve2018, McPherson1987, Ferray1988, Gariepy2014, Geneaux2016, Kong2017, Gauthier2016,Dorney2019}. 
Here we report on HHG driven by a planar polarization Möbius strip with GAM charge $j_\gamma^{\text{IR}}=\frac{1}{2}$, which is formed by superimposing two infrared (IR) driving beams carrying different OAM and SAM. We show that the GAM charge scales linearly with the harmonic order, yielding a half-integer angular momentum spectrum.  

A major challenge faced in this program is the characterization of the OAM carried by the extreme-ultraviolet (XUV) harmonics. The techniques commonly used in the visible domain, based on interference or mode-conversion with tilted lenses, are difficult or impossible to implement here, due to the absence of transmissive optics at XUV wavelengths. We instead propose two novel methods. The first is based on spatial interference between two harmonic vortices, and allows to retrieve their relative phase variation. The second uses transverse mode conversion by a spherical mirror \cite{Allen1992, ONeil2000, Padgett1999}, which had not been achieved in the XUV domain so far. Both techniques might be promising in future experiments involving XUV beams with angular momentum.

\begin{figure*}
    \centering
    \includegraphics[width=15cm]{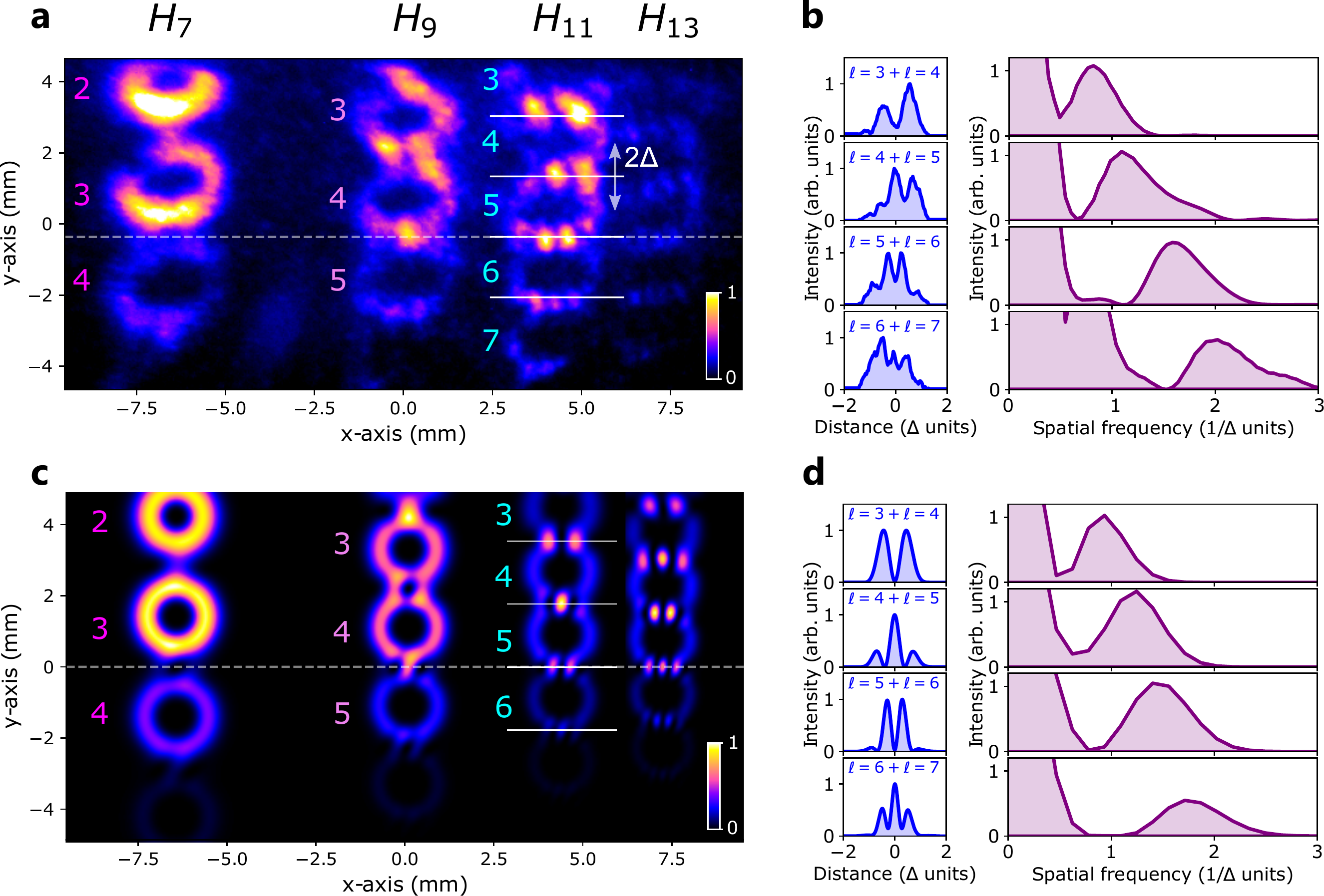}
    \caption{\textbf{High harmonic spectrum generated by a combination of $\ell_1=0$ and $\ell_2 = 1$ linearly polarized non-collinear driving beams}. \textbf{(a)} Experimental intensity profiles of harmonic $7$, $9$, $11$ and $13$.  The colored digits indicate the OAM charge of the nearest XUV vortex. The bisector of the driving beams is indicated with a horizontal dashed line. $\Delta$ is half the distance between the centers of two neighboring vortices, which only depends on the harmonic order and the relative angle $\theta$ of the driving beams. \textbf{(b)} Line-outs of harmonic 11 along the white horizontal lines in (a) (left), and corresponding spatial Fourier transform (right), with spatial frequencies in units of $1/\Delta$. 
   \textbf{(c)} Simulated intensity profiles of harmonic $7$, $9$, $11$ and $13$.  \textbf{(d)} Line-outs of harmonic 11 along the white horizontal lines in (c) (left), and corresponding spatial Fourier transform (right).}
    \label{fig:fig2}
\end{figure*}

We use a high intensity IR Möbius strip field to drive HHG in a jet of argon gas. This field is obtained by overlapping, temporally and spatially, the pulses from two 800 nm wavelength laser beams (refered to as beam 1 and 2) carrying different OAM charges $\ell_1 = 0$ and $\ell_2 = 1$, and opposite helicity (i.e. they have orthogonal circular polarization) $\sigma_1 = +1$ and $\sigma_2 = -1$ \cite{Galvez2017}. 
The associated coordination factor $\gamma$ is a half-integer \cite{Ballantine2016}:
\begin{equation}
\label{eq:gam charge}
    \gamma = \frac{\ell_2 - \ell_1}{2} = \frac{1}{2},
\end{equation}
and both driving beams carry the same GAM charge: $j_\gamma^{\text{IR}} = \ell_1 + \gamma \sigma_1 = \ell_2 + \gamma \sigma_2 = \frac{1}{2}$. Thus, the total IR field is in a well-defined GAM state. Fig.\,\ref{fig:fig1}.a shows the IR polarization state in the transverse plane, when the two driving beams are collinear. On the optical axis, the polarization is purely circular, forming  a "C-point" \cite{Nye1983, Berry2004}. Around the C-point, the field is elliptically polarized and the orientation of the ellipse varies by $\pi$ when travelling along a loop about the optical axis (Fig.\,\ref{fig:fig1}.a').
The winding number associated to the C-point is thus $I_C = \frac{1}{2}$ \cite{Freund2009}. 
In order to visualize the Möbius strip topology of the field, the polarization ellipses computed along the grey dashed circle in Fig.\,\ref{fig:fig1}.a are stacked, 
resulting in a twisted strip traced by the major axis of the ellipses, the two ends of which corresponds to the azimuth $0$ and $2\pi$. Connecting them, we obtain a Möbius strip 
whose single edge is traced by the tips of the ellipses (Fig.\,\ref{fig:fig1}.b) 
\footnote{The IR field considered here is not strictly speaking a three dimensional Möbius strip. In the work of Bauer \textit{et al.} \cite{Bauer2016, Bauer2017}, Galvez \textit{et al.} \cite{Galvez2017} and Freund \textit{et al.} \cite{Freund2005, Freund2009}, non-paraxial fields are studied, where the 3D orientation of the polarization ellipse major axis draws Möbius strips in real space. On the other hand, we assume here that the paraxial approximation is valid, i.e. the field's polarization is confined in the plane transverse to the propagation direction. Thus, the polarization of the IR field is only topologically equivalent to a Möbius strip, and does not yield real three dimensional objects. In this way, our approach is the same as that of Pisanty \textit{et al.} \cite{Pisanty2019Nature}, where a torus knot topology is arising from paraxial beams.}.
\begin{figure*}
    \centering
    \includegraphics[width=15cm]{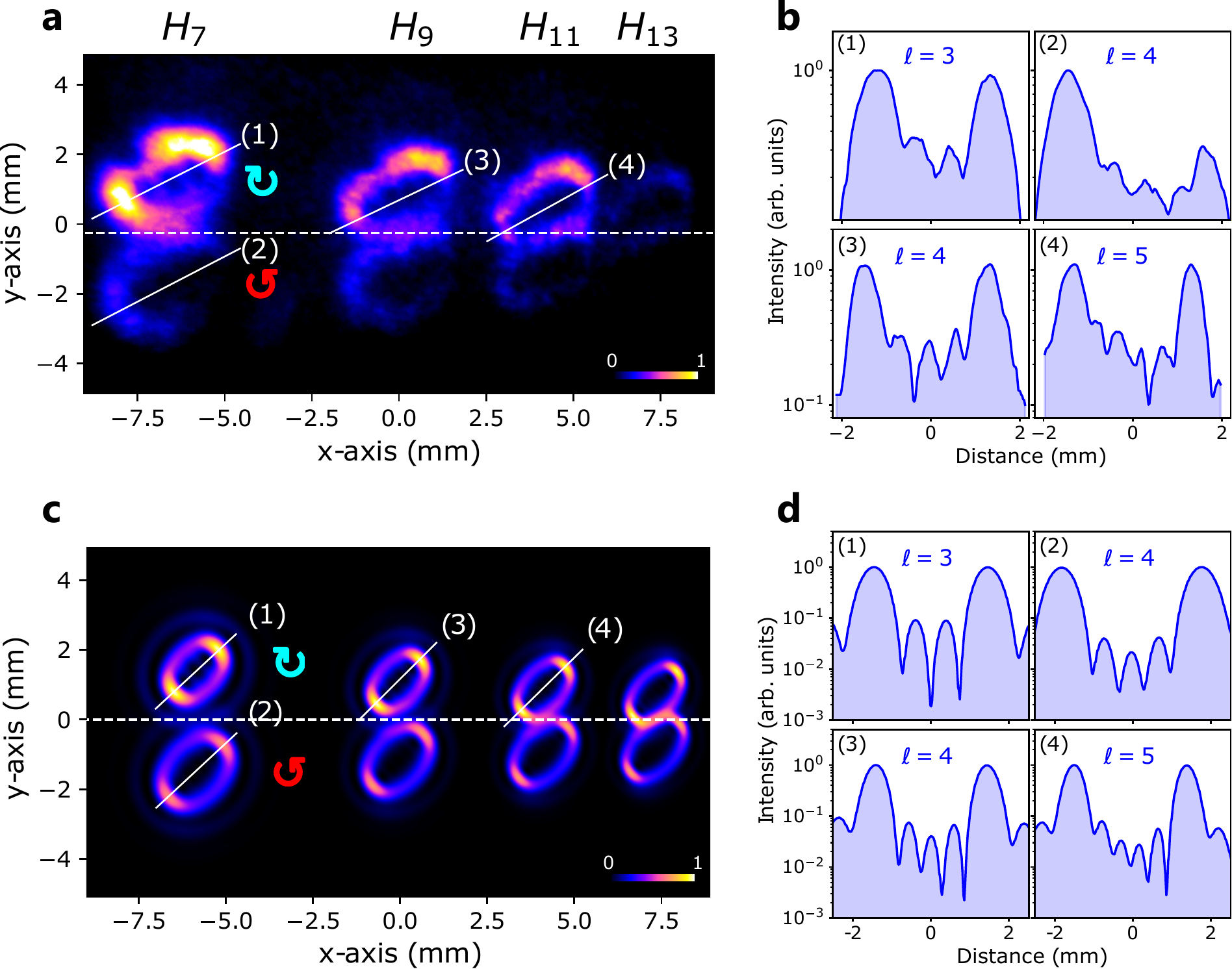}
    \caption{\textbf{High harmonic spectrum obtained with a $I_C=\frac{1}{2}$ polarization Möbius strip as the driving field}. \textbf{(a)} Experimental intensity profiles of harmonic $7$, $9$, $11$ and $13$. The bisector of the driving beams is indicated with a horizontal dashed line. Circular arrows indicate the helicity of the light above and below the bisector (blue: CR,"$+$" emission channel; red: CL,"$-$" emission channel). The two beamlets of each harmonic order have unequal intensity because the two driving beams have unequal pulse energy. \textbf{(b)} Line-outs of the XUV intensity along the corresponding lines in (a), in log scale. The number of intensity minima corresponds to the OAM charge of the vortex considered. \textbf{(c)} Simulated intensity profiles of harmonic $7$, $9$, $11$ and $13$. \textbf{(d)} Line-outs of the XUV intensity along the corresponding lines in (c), in log scale.}
    \label{fig:fig3}
\end{figure*}
The interaction of this IR field with the atomic gas results in the emission of high-order odd harmonics. Since the efficiency of the process decreases exponentially with the ellipticity of the driving field \cite{Budil1993}, XUV emission mostly occurs along the circle where beams 1 and 2 have equal intensity, i.e. where the IR polarization is linear (Fig.\,\ref{fig:fig1}.a and Supplementary B).

We model HHG using the so-called thin-slab model (Methods). The results are diplayed in
Fig.\,\ref{fig:fig1}.c. Each harmonic is itself a polarization Möbius strip (top panel), i.e. is a superposition of OAM and SAM states. This can be understood within a parametric picture of HHG.
In order to emit the $q\textsuperscript{th}$ harmonic, with $q$ odd, $p$ photons can be taken from beam 1, and $q-p$ photons from beam 2. 
OAM conservation \cite{Bloembergen1980, Hernandez-Garcia2013, Gariepy2014, Geneaux2016,  Kong2017, Gauthier2016} implies that the resulting emission channel $(q,p)$ carries the OAM $\ell_{(q, p)} = p\ell_1 + (q-p)\ell_2 = q-p$. SAM conservation forbids all photonic channels yielding a XUV photon with spin $\sigma_{(q, p)} \neq \pm 1$ \cite{Fleisher2014, Pisanty2014}. Thus, all harmonics are expected to be superpositions of two emission channels, that we refer to as "$+$" and "$-$", carrying the SAM $\sigma_{\pm}^{(q)} = \pm 1$ and the  OAM $ \ell_{\pm}^{(q)} = (q\mp1)/2$ \cite{Turpin2017}.
Such a photon picture also predicts that the GAM charge of harmonic $q$ should be 
$j_\gamma^{(q)} = \ell_{+}^{(q)} + \gamma \sigma_{+}^{(q)} = \ell_{-}^{(q)} + \gamma \sigma_{-}^{(q)} = q/2$. In the time domain, the GAM charge corresponds to invariance under a dynamical symmetry consisting of a coordinated rotation and a time delay \cite{Alon1998}. Inspecting the complex XUV field $\textbf{E}$, this linear increase of the GAM with the harmonic order appears when we plot the so-called rectifying phase, defined as $\Phi = \frac{1}{2} \text{arg} (\textbf{E}\cdot \textbf{E})$ \cite{Dennis2002}. It shows $2 \times j_\gamma^{(q)}$ phase jumps of $\pi$  (bottom panel of Fig.\,\ref{fig:fig1}.c).

To experimentally verify these predictions, one has to access the SAM and OAM content of the high harmonics. The spatially-resolved measurement of the polarization and OAM-state of those composite beams 
would be extremely challenging, due to the cumbersome operation of phase retarders \cite{Vodungbo2011, Fleisher2014, Siegrist2019}, and Hartman sensors \cite{Gauthier2016, Sanson2018} in the XUV. Here, we instead lift the degeneracy of the "$+$" and "$-$" emission channels by introducing a small angle $\theta$ (50 mrad) between the two driving IR beams. This way, the two counter-rotating channels are separated angularly, as shown in Fig.\,\ref{fig:fig1}.f, as a consequence of the conservation of the photon linear momentum \cite{Bertrand2011, Hickstein2015}. 
Because of the non-collinear crossing of the drivers, the IR Möbius strip now exhibits several right and left-handed twists (Fig.\,\ref{fig:fig1}.d-e and Supplementary Video). Still, the net number of twists remains one, i.e. one can untwist the Möbius strip of Fig.\,\ref{fig:fig1}.e to obtain that of Fig.\,\ref{fig:fig1}.b, as a consequence of the topological protection of the winding number \cite{Freund2009}.

In the first step of the experiment, we set the polarization of the two superimposed IR beams to linear vertical, while keeping their OAM equal to $\ell_1=0$ and $\ell_2=1$. In this configuration, which is not constrained by SAM conservation, all HHG photon channels $(q, p)$  are emitted. This process corresponds to HHG driven by a fork-shaped intensity and phase grating, making it possible to draw an analogy with holography \cite{Kong2017}. 
The measured XUV spectrum is shown in Fig.\,\ref{fig:fig2}.a. The non-collinear angle $\theta$ was adjusted so that, for harmonic 11, the bottom part of the beamlet $(q, p)$ spatially overlaps with the top part of the following one, i.e. $(q, p+1)$. In the overlapping region, the spatial phases of the two neighbouring vortices vary in opposite directions and an interference pattern is observed (Fig.\,\ref{fig:fig2}.b). As we consider higher values of $p$, the OAM gets higher, thus the spatial period of the interference pattern gets smaller. Comparing with simulation results (Fig.\,\ref{fig:fig2}.c,d), the periodicity of the interference fringes provides a measurement of the absolute value of the algebraic sum of the two OAM charges (Supplementary E,F). The result is compatible with the predicted law $\ell_{(q, p)} = q-p$. Note that the interference pattern is also clearly visible for harmonics 9 and 13 (Supplementary G). Incidentally, this experiment allows to observe the interference between two distinct photonic channels of a given harmonic. In particular, it could provide a way to retrieve their relative phase which, to the best of our knowledge, was not achieved thus far. This unique capability is provided by the azimutal phase variation and the high divergence of OAM beams.

\begin{figure}
    \centering
    \includegraphics[width=7cm]{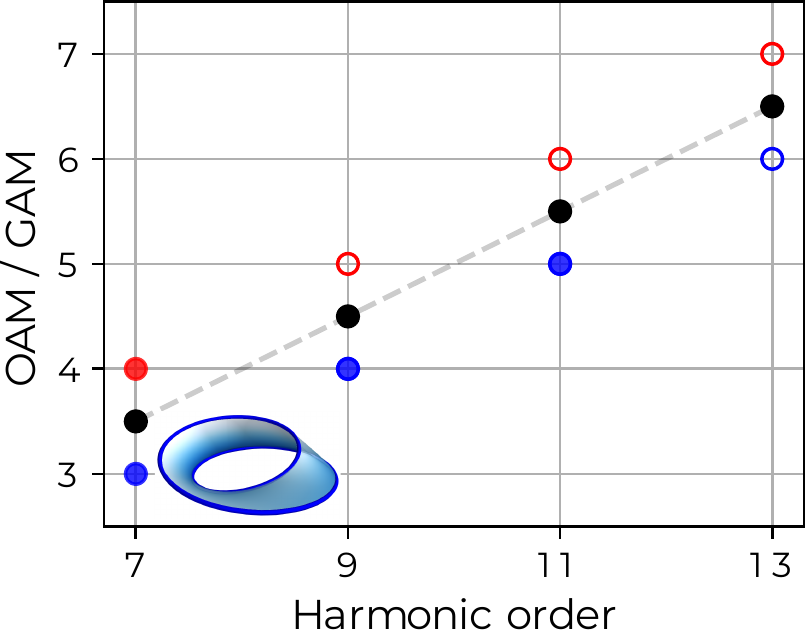}
    \caption{\textbf{Conservation of the generalized angular momentum.} Each harmonic order $q$ is in a superposition of two modes of OAM $(q-1)/2$ and $(q+1)/2$, with SAM $+1$ (blue circles) and $-1$ (red circles), respectively. The GAM charge, defined as $j_{\frac{1}{2}} = \ell + \frac{1}{2} \sigma$, is shown with black dots. The red and blue dots that are filled correspond the the OAM measured by the LG to HG conversion, while the empty ones show the measurement based on XUV vortices interference (performed in linearly polarized light). The inset 3D surface illustrates the Möbius strip topology of the harmonics generated in collinear geometry. }
    \label{fig:GAM}
\end{figure}

We then set the polarization of the driving beams to counter-rotating circular, and generate high harmonics with the resulting polarization Möbius strip. We obtain the harmonic spectrum shown in Fig.\,\ref{fig:fig3}.a. As expected, all beamlets are completely extinguished, except for the two located about the bisector of the drivers, for all harmonic orders. Comparing with Fig.\,\ref{fig:fig2}.a, we also observe complete suppression of the interference fringes in the overlapping region of the vortices. These two facts indicate that the remaining beamlets have opposite $\pm 1$ SAM, inherited from the driving IR beams, and that the interference pattern now takes the form of a complex rotating polarization direction in the region where they overlap.
As a downside, it is now impossible to quantify the OAM content with the previous method. Instead, we take advantage of controllable astigmatism of the re-focusing optics. By increasing the incidence angle of the XUV light on a spherical mirror (Methods) from normal up to 15\textdegree, we obtain a partial Laguerre-Gaussian to Hermite-Gaussian mode converter \cite{Allen1992, ONeil2000, Padgett1999}
(Supplementary I,J).  Counting the number of intensity minima on the diagonal of each harmonic vortex \cite{Vaity2013} (Fig.\,\ref{fig:fig3}.b,d), we find that the beamlets $(q, (q\pm1)/2)$ carry the OAM $(q\mp1)/2)$
, in agreement with the OAM value measured with linear polarization, and the prediction of the photon counting method. 
With the values of both the OAM and SAM of the high harmonics vortices in hand, we can  now compute their GAM charge.  
According to the experimental and theoretical results, the '+' and '-' beamlets carry the OAM $\ell_{\pm}^{(q)} = (q\mp 1)/2$ and the SAM $\sigma_{\pm}^{(q)} = \pm1$. Thus, despite having different OAM charges and helicities, they both carry the same GAM charge
\begin{equation}
    j_{\gamma}^{(q)} =  \ell_{\pm}^{(q)}+\frac{1}{2}\sigma_{\pm}^{(q)} =  \frac{q}{2}.
\end{equation}
The GAM charge is therefore a property of harmonic $q$ that does not depend on the number of photons absorbed from each driving IR beam. It is equal to $q$ times that of the fundamental beams 
\begin{equation}
      j_\gamma^{(q)} = q \times j_\gamma^{\text{IR}},
\end{equation}
yielding $j_\gamma^{(7)} =7/2$, $j_\gamma^{(9)} =9/2$, $j_\gamma^{(11)} =11/2$, and so on. This trend is shown in Fig.\,\ref{fig:GAM}. In collinear driving geometry, where the '+' and '-' beamlets are both emitted on axis, the GAM charge varies linearly with the harmonic order, while the OAM splits into two components of opposite SAM. 

If the absence of a spectrometer in the setup, the XUV emission depicted in Fig.\,\ref{fig:fig3} would simply consist of two broadband XUV rings separated angularly, with opposite helicity, each ring being a comb of harmonics with increasing OAM values. Thus, in the temporal domain, this non-collinear scheme allows the simultaneous generation of right and left circularly polarized single-helix attosecond light springs \cite{Hernandez-Garcia2013, Geneaux2016, Turpin2017} (Supplementary K). If the two driving beams are collinear, one obtains a light spring with Möbius strip topology \cite{Turpin2017}.

In conclusion, we demonstrated experimentally that the GAM charge $j_\gamma$ is conserved in nonlinear optics and that the topology of the fundamental field is transferred to the generated light.
$j_\gamma$ is thus the appropriate quantum number to describe nonlinear phenomena involving light fields invariant under coordinated rotations.  
Incidentally, we directly observed the interference between two photon channels contributing to the same harmonic order. Our scheme could thus contribute to the validation of photon-based theories of HHG \cite{Li2018}. By exploiting transverse mode conversion by astigmatic focusing, we also introduced an OAM measurement method in the XUV that is of general usefulness. It may serve the ever growing interest in applications of focused harmonic beams carrying OAM, SAM, or both, for instance in angular-momentum-induced dichroisms \cite{ Zambrana2014, Forbes2021, Fanciulli2022}. 
The production of XUV light beams with Möbius strip topological singularities, beyond answering fundamental questions about the conservation of GAM, also opens new perspectives in ultrafast light-matter interaction. In particular, whether the topology of light could be modified upon interaction with matter remains a pending question. Our approach could be extended to light beams with time-varying GAM, along the lines of Ref.\,\cite{Rego2019}, bichromatic fields \cite{Kong2019, Dorney2019, Paufler2018}, HHG driven by single-cycle pulses \cite{Huang2018}, or tightly-focused fields \cite{Fang2021}; further expanding the capabilities of ultrafast angular momentum-based spectroscopy. Finally, our results, combined with the recent progress of the quantum optical description of HHG \cite{Lewenstein2021}, could enlighten the discussion triggered by Ballantine \textit{et al.} \cite{Ballantine2016} about the photon statistics of light beams with non-integer GAM.\\

\textbf{Methods}\\

\textbf{Experimental setup}
The fundamental laser is a Ti:Saphire regenerative amplifier, delivering 25 fs pulses centered at a 800 nm wavelength, with a 1 kHz repetition rate and a 2 mJ pulse energy. The laser is split in two beams (beams 1 and 2) by an intensity beam splitter, and propagates in a Mach-Zender interferometer (Supplementary A). A spiral phase plate (SPP), made of 16 azimuthal steps and manufactured by SILIOS
Technologies, gives an OAM charge $\ell_2 = 1$ to beam 2 (Supplementary D). The SPP is designed to imprint an optical path varying from $0$ to $\lambda_{IR} = 800$ nm over a $2\pi$ azimuthal range. At the output of the interferometer, the two beams are recombined non-collinearly. Their relative angle $\theta$ is kept sufficently small (50 mrad) in order to remain in the paraxial regime and avoid any spin-orbit coupling of light \cite{Bliokh2015}. A zero-order half waveplate (HWP) is placed in beam 2, and a zero-order quarter waveplate (QWP) is common to both beams after the interferometer. In order to obtain counter-rotating circular IR beams, the slow axis of the HWP is precisely aligned at 45\textdegree, so that the field of beam 2 becomes horizontally polarized, and the axis of the QWP common to both beams is precisely set at 45\textdegree (Supplementary H). Each beam (collimated, with a diameter of about 14 mm) is focused using a plano-convex lens with 75 cm focal length. The diameter of the focal spot size is about 50 $\mu m$. After passing though all the optical elements, a pulse of beam 1 (respectively beam 2) has an energy of 0.61 mJ (respectively 0.52 mJ). The optical path difference between the interferometer's arms is set to zero by means of a delay stage placed in beam 1. HHG occurs in an effusive jet of argon atoms issued from a 500 $\mu$m nozzle, located  in the focal plane of both beams.
In order to observe the inter-vortex interference, or to achieve controllable LG to HG mode conversion, astigmatism must be limited, which prevents the use of cylindrical mirrors and gratings with variable steps that are commonly employed in HHG experiments. We use instead a gold grating with uniform periodicity (600 grooves per mm). The XUV spot size on the grating must be minimized, in order to limit dispersion. The harmonics are thus focused by a near normal incidence spherical mirror with boron carbide (B$_4$C) coating and 600 mm curvature radius. The angle of incidence on the mirror allows us to control the astigmatism as per requirement. The near normal incidence reflections on the sherical mirror and the grating prevent efficient detection of wavelength shorter than 50 nm, which approximately corresponds to harmonic order $15$ (Supplementary A). The harmonics propagate to the far field where a set of micro-channel plates and a phosphor screen is located. Finally, the XUV spectrum at the back of the phosphor screen is recorded with a CMOS camera (Hamamatsu Photonics, model ORCA Spark C11440-36U). The experimental images shown in this work were acquired by averaging 30 frames, with 0.6 s exposure time for each frame.\\

\textbf{HHG simulation} 
The numerical model (the so-called "thin-slab model" \cite{Rego2017}) assumes that the gas jet is infinitely thin in the longitudinal direction, so that HHG occurs in a 2D transverse plane. Thus we discard any longitudinal phase matching effect. 
This gas sheet is placed at the crossing point of the two driving beams, where their respective focii are located. The intensity of the HHG emission in the focal plane is computed as $I_{\text{IR}}^{q_{\text{eff}}} e^{-(\epsilon_{\text{IR}}/\epsilon_0)^2}$, where $I_{\text{IR}}$ is the local intensity of the IR field (with a peak intensity of $10^{14}$ W cm$^{-2}$), $q_{\text{eff}} = 3.5$ is a typical effective order of nonlinearity \cite{Rego2017}, and $\epsilon_{\text{IR}}$ and  $\epsilon_0 = 0.2$ are the local ellipticity of the IR and a threshold ellipticity, respectively \cite{Budil1993}.
The phase of the harmonic order $q$ at the focus is thus given by $q \phi_{\text{IR}} + \alpha I_{\text{IR}}$, with $\phi_{\text{IR}}$ the local phase of the infrared field and $\alpha I_{\text{IR}}$ the atomic phase, proportional to the IR intensity. The strong field parameter $\alpha = 2 \times 10^{-14}$ cm$^2$ W$^{-1}$ corresponds to short electronic trajectories.  The polarization state of the generated harmonics at focus is assumed to exactly match that of the infrared. The resulting XUV field is then propagated to the far field using the Fraunhofer propagation operator \cite{Rego2017}. In the simulation, the first beam is an ideal right circularly polarized Gaussian beam, and the second beam is an ideal left circularly polarized Laguerre-Gaussian mode of indices $(\ell=1, \rho = 0)$, both having the same waist radius $w_0 = 50$ $\mu$m. The peak intensity of the LG beam is 0.85 times that of the Gaussian beam. The non-collinear angle $\theta$ is adjusted so that, in the far field, the distance $\Delta$ between two XUV vortices matches with the experiment (i.e., we did not simulate the re-focusing optics). This configuration is achieved for $\theta = 40$ mrad in the simulation. Quantum simulations based on the strong field approximation (SFA) were also performed to corroborate the temporal emission of the Möbius strip driven harmonics in the form of circularly polarized attosecond light springs (Supplementary K).

\begin{acknowledgments}

This work was supported by the French “Investments for the Future” of the Agence
Nationale pour la Recherche (Contracts No. 11-EQPX0005-ATTOLAB, ANR-14-CE32-0010 - Xstase, ANR-10-LABX-0039-PALM ATTO-Foam and ANR HELIMAG ANR-21-CE30-0037), the Scientific Cooperation Foundation of Paris-Saclay University through the funding of the OPT2X research project (Lidex 2014), by the Île-de-France region through the Pulse-X project and by the European Union’s Horizon 2020 Research and Innovation Programme No. EU-H2020-LASERLAB-EUROPE-654148.
E.P. acknowledges support by the Royal Society through the University Research Fellowship UR\textbackslash R1\textbackslash211390.
C.H.-G. acknowledges support from the European Research Council (ERC) under the European Union’s Horizon 2020 Research and Innovation Program (Grant Agreement No. 851201), Ministerio de Ciencia e Innovación (RYC-2017-22745, PID2019-106910GB-I00), and Junta de Castilla y León (SA287P18).
A.Z.K acknowledges  financial support from the Brazilian Agencies, Conselho Nacional de Desenvolvimento Tecnológico (CNPq), Fundação Carlos Chagas Filho de Amparo à Pesquisa do Estado do Rio de Janeiro (FAPERJ), Coordenação de Aperfeiçoamento de Pessoal de Nível Superior (CAPES – Finance Code 001), the Brazilian National Institute of Science and Technology of Quantum Information (INCT-IQ 465469/2014-0), and São Paulo Research Foundation (FAPESP), grant \# 2021/06823-5.

\end{acknowledgments}

\section*{Authors contributions}

M.L, M.V and T.R conceived the experiment. M.L, M.V and J-F.H performed the experiment and collected the data. M.L analyzed the data. M.G and C.H-G performed the simulations. M.L, T.R, E.P, C.H-G, A.Z-K and M.G interpreted the results. M.L and T.R wrote the manuscript with contributions from all authors.

\section*{Competing interests}
The authors declare no competing interests.

\bibliography{apssamp}

\end{document}


\preprint{APS/123-QED}

\title{\Large SUPPLEMENTARY INFORMATION:\\ Conservation of a Half-Integer Angular Momentum in Nonlinear Optics with a Polarization Möbius Strip}

\author{Martin Luttmann}\email{martin.luttmann@cea.fr}
\affiliation{%
Universit\'e Paris-Saclay, CEA, CNRS, LIDYL, 91191 Gif-sur-Yvette, France
}

\author{Mekha Vimal}%
\affiliation{%
Universit\'e Paris-Saclay, CEA, CNRS, LIDYL, 91191 Gif-sur-Yvette, France
}

\author{Matthieu Guer}
\affiliation{%
Universit\'e Paris-Saclay, CEA, CNRS, LIDYL, 91191 Gif-sur-Yvette, France
}
\affiliation{Grupo de Investigación en Aplicaciones del Láser y Fotónica, Departamento de Física Aplicada, University of Salamanca, Salamanca E-37008, Spain}

\author{Jean-François Hergott}
\affiliation{%
Universit\'e Paris-Saclay, CEA, CNRS, LIDYL, 91191 Gif-sur-Yvette, France
}

\author{Antonio Z. Khoury}
\affiliation{Instituto de Física, Universidade Federal Fluminense, 24210-346 Niterói, RJ, Brazil}


\author{Carlos Hern\'andez-Garc\'ia}
\affiliation{Grupo de Investigación en Aplicaciones del Láser y Fotónica, Departamento de Física Aplicada, University of Salamanca, Salamanca E-37008, Spain}

\author{Emilio Pisanty}
\affiliation{Department of Physics, King's College London, Strand Campus, WC2R 2LS, London, UK}

\author{Thierry Ruchon}
\email{thierry.ruchon@cea.fr}
\affiliation{%
Universit\'e Paris-Saclay, CEA, CNRS, LIDYL, 91191 Gif-sur-Yvette, France
}

\date{\today}

\maketitle




\subsection{Experimental setup}

\begin{figure*}
    \centering
    \includegraphics[width=13cm]{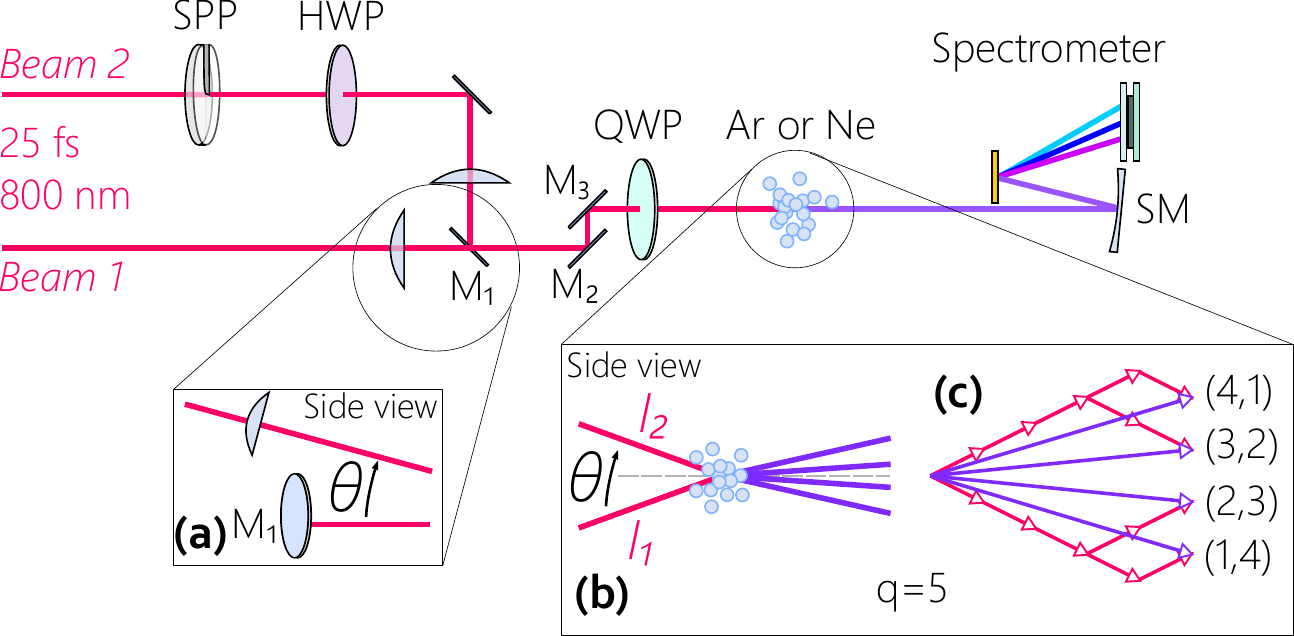}
    \caption{Experimental setup for driving HHG with an infrared polarization Möbius strip beam. The main pannel shows the setup as seen from above. (a, b) are side views of specific parts of the setup. SPP: spiral phase plate, HWP: half waveplate, QWP: quarter waveplate, SM: spherical mirror. Not represented: attenuator in beam 2.}
    \label{fig:setup}
\end{figure*}

\begin{figure*}
    \centering
    \includegraphics[width=13cm]{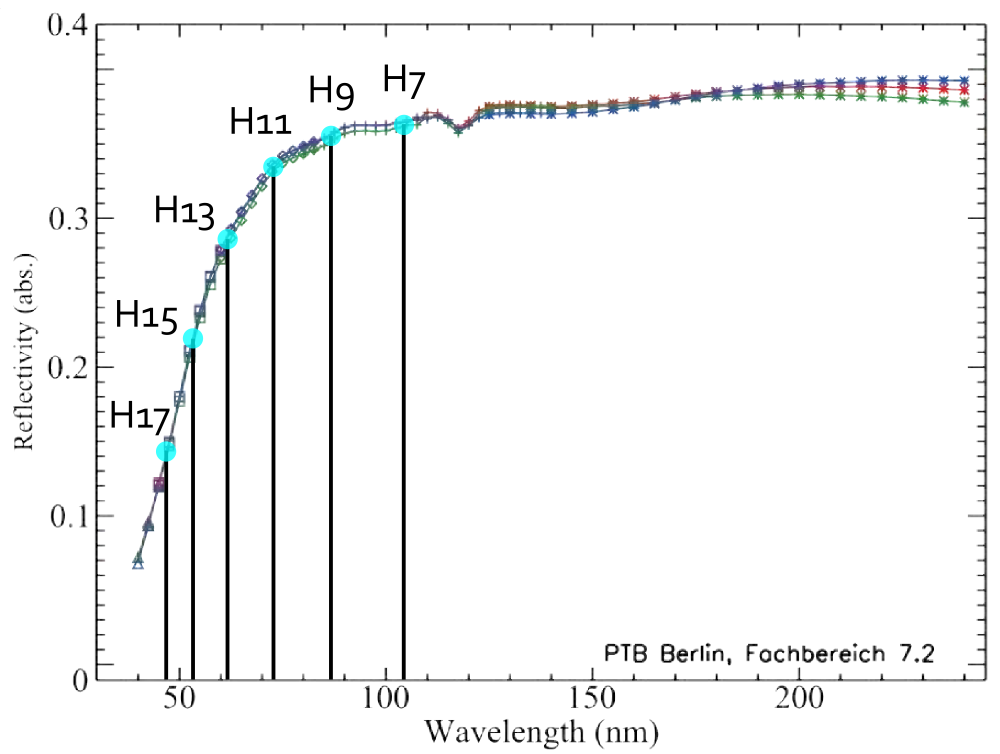}
    \caption{Reflectivity of the B$_4$C mirror as a function of the wavelength. Data from Fraunhofer IOF, reproduced in Ref.\,\cite{Ruf2012}, fused silica substrate coated by 16 nm B4C.}
    \label{fig:Hartmut}
\end{figure*}

The experimental scheme is shown in Fig. SM\,\ref{fig:setup}. An infrared input beam (Ti:Sapphire laser, 1kHz repetition rate, 2 mJ energy per pulse, 25 fs full width half maximum duration) is split into two arms by an intensity beam splitter (not shown in the figure). A spiral phase plate (SPP) is used to impart an OAM $\ell_2 = 1$ to \textit{beam 2}. A set of half wave-plate (HWP) and quarter wave-plate (QWP) allows us to obtain counter-rotating circularly polarized drivers. The beams from the two arms are re-combined non-collinearly - \textit{beam 1} propagates above the last mirror ($M_1$ , inset (a)) - and focused into a gas jet of Ar atoms. $M_2$ and $M_3$ are aligned so that the bisector of the drivers (dashed line in inset (b)) is parallel to the optical axis. Because of the non-collinear geometry, a given harmonic splits into several beamlets carrying different OAM and SAM. From a photon point of view, each beamlet corresponds to a specific combination $(p, q-p)$ of photons absorbed from each driving beam (inset (c), with $q=5$). The total intensity of \textit{beam 2} can be adjusted by the mean of an additional HWP and a polarizer (not shown). Finally, the XUV light is re-focused by a spherical mirror (SM) with 300 mm focal length, diffracted by a grating and observed with a set of micro-channel plates and phosphore screen. The near-normal incidence of the XUV light on the SM allows to control the astigmatism and eventually the spatial profiles of the harmonics, at the cost of a reduced reflectivity and bandpass. In practice, harmonics above $q=15$ could not be detected  (Fig. SM\,\ref{fig:Hartmut}).

\subsection{Numerical analysis of the IR polarization Möbius strip}

\begin{figure*}
    \centering
    \includegraphics[width=13cm]{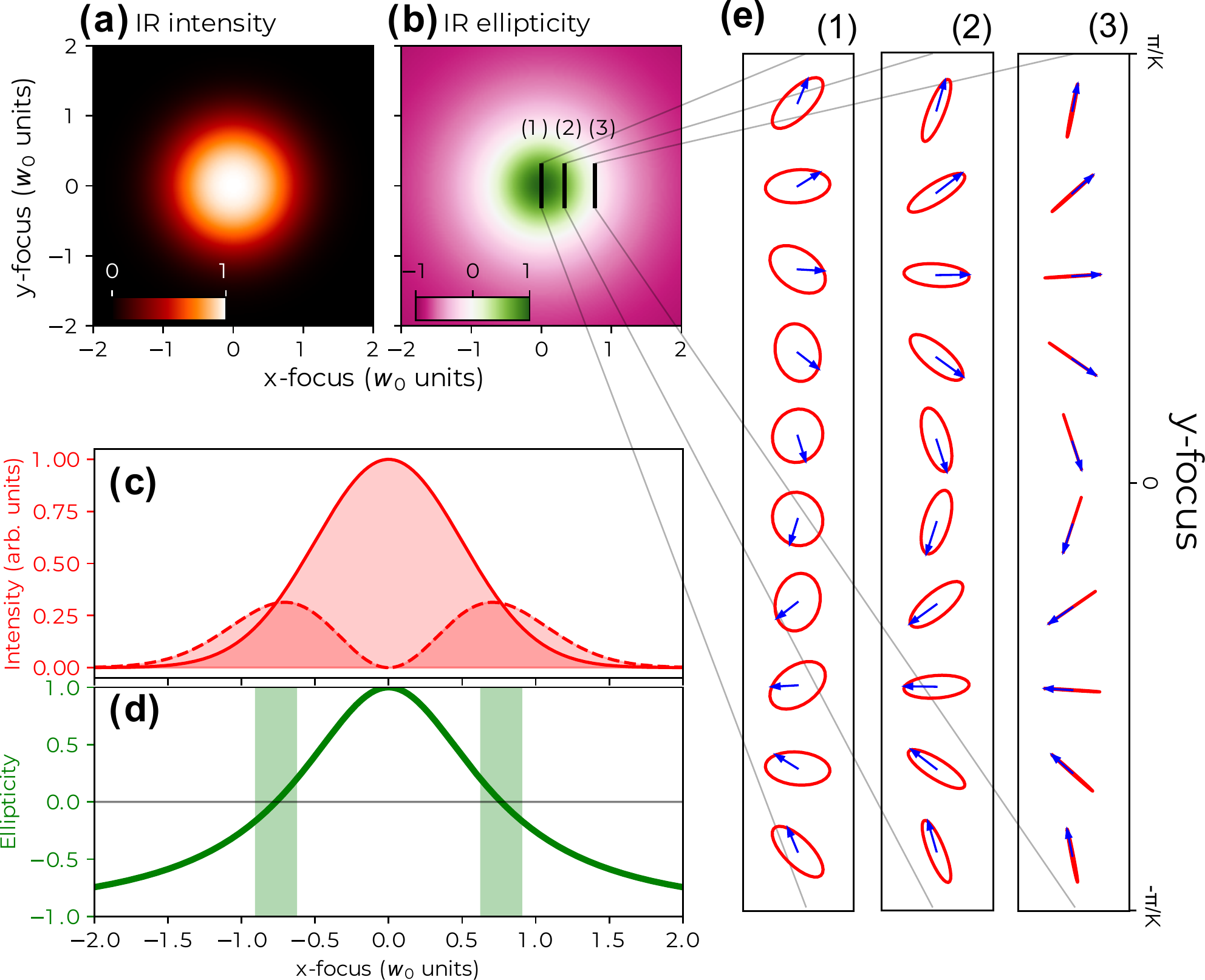}
    \caption{Features of the IR polarization Möbius strip, computed with theoretical ideal beams. \textbf{a} Total intensity at focus. \textbf{b} Ellipticity, defined as $\frac{I_1-I_2}{I_1+I_2}$, with $I_1$ and $I_2$ the intensity of beam 1 and 2 respectively. \textbf{c} Line-out of the intensity of beam 1 (continuous line) and beam 2 (dashed line) along the line $y=0$. \textbf{d} Line-out of the ellipticity along the line $y=0$. Green vertical zones indicate the regions where HHG occurs efficiently. \textbf{e} Polarization state of the IR field, along 3 lines of length $2\pi/K$, with $K=2\pi \sin \theta/\lambda_{\text{IR}}$, i.e. one period of the ellipse orientation grating.}
    \label{fig:IRfield numerical}
\end{figure*}

Because of the Gaussian and LG transverse profiles of the two fundamental beams, there is always a radius $r_0$ at which they have the same local intensity, regardless of their relative total intensity (Fig. SM\,\ref{fig:IRfield numerical}). Denoting $\alpha$ the energy ratio of the beams, one finds $   r_0 = \sqrt{\frac{w_0}{2\alpha}}$, where $w_0$ is the driving beams waist. With our experimental parameters, we have $r_0 \approx 39$ microns. High harmonics are thus mainly emitted along the circle of radius $r_0$.

\subsection{Möbius strip video}

This movie shows the three-dimensional representation of the IR polarization Möbius strip (Fig.\,1.A, D) of the main text) in a rotating frame, for the collinear ($\theta = 0$ mrad) and the non-collinear scheme ($\theta = 50$ mrad). The blue ellipse is a cross-section of the Möbius strip in a plane that rotates alongside the camera. The red curve shows half of the edge of the strip. This movie allows to count the number of $\pi$ rotations of the polarization ellipse as the azimuthal coordinate varies. The collinear Möbius strip contains a single counter-clockwise twist, while the non-collinear Möbius strip contains 3 counter-clockwise and 2 clockwise twists.

\subsection{Measurement of the topological charge carried by beam 2}

\begin{figure}
\centering
\begin{minipage}{.5\textwidth}
  \centering
  \includegraphics[width=.6\linewidth]{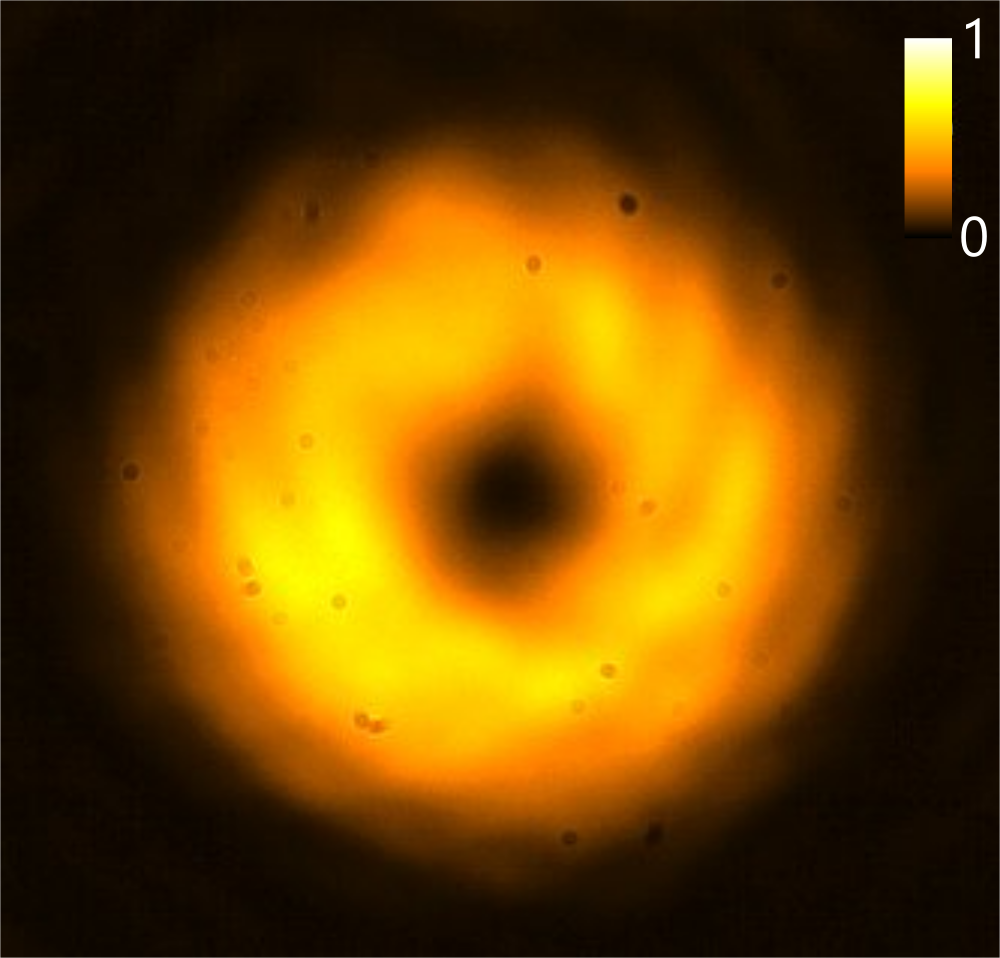}
  \caption{Intensity profile of beam 2 at focus.}
  \label{fig:LG focus}
\end{minipage}%
\begin{minipage}{.5\textwidth}
  \centering
  \includegraphics[width=.6\linewidth]{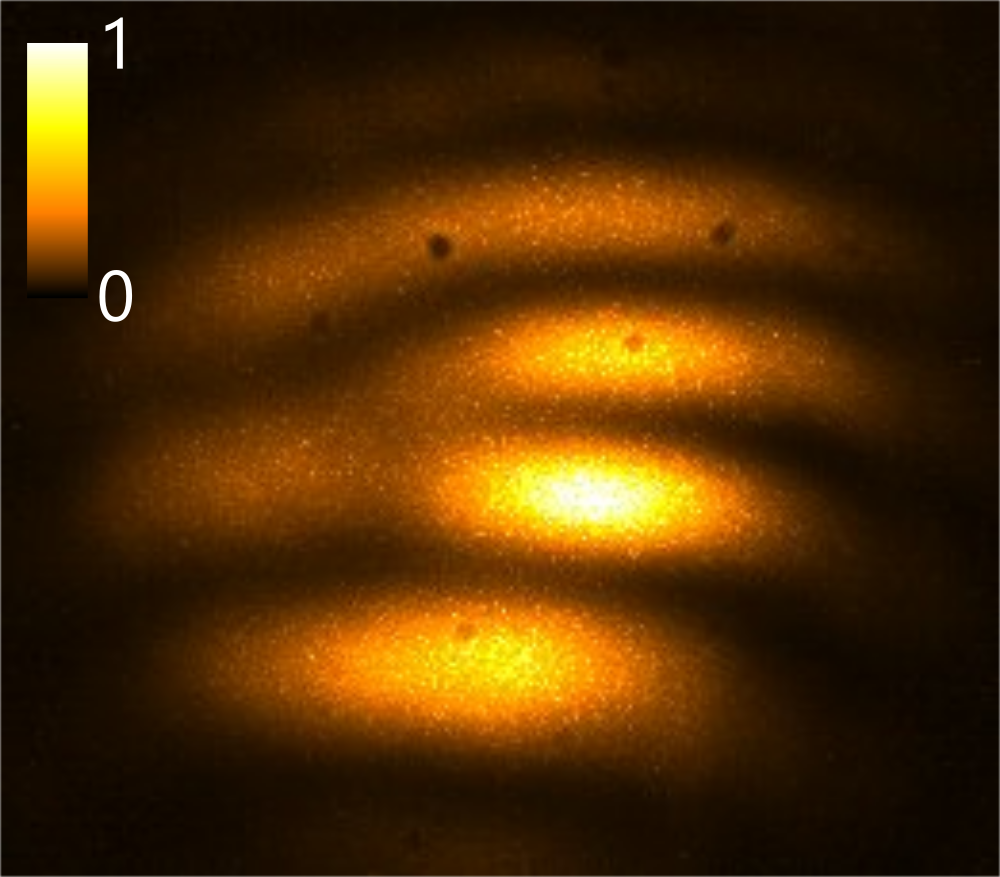}
  \caption{Interference pattern produced by intersecting beam 1 (gaussian beam) and beam 2, with a relative angle of 25 mrad.}
  \label{fig:OAM measurement}
\end{minipage}
\end{figure}

The quality of the OAM mode in beam 2 is of prime importance when driving HHG, since uneven intensities and wavefronts of the IR profile are enhanced by the non-linearity of the process. A CCD camera was placed at the focus of the IR beams, with a microscope objective. Fig. SM\,\ref{fig:LG focus} is an image of the focus of beam 2. The beam profile has an annular profile showing excellent rotationnal symmetry about the optical axis. In order to measure the vortex topological charge, we record an image of the focus when beam 1 (gaussian beam) and beam 2 are superimposed non-collinearly, both having vertical linear polarization (corresponding to S polarization on all reflective surfaces of our setup, Fig. SM\,\ref{fig:OAM measurement}). The interference pattern shows a "fork shape", typical of the interference between plane and helical waves. The topological charge of the vortex is given by the difference in the number of fringes on each side of the dislocation \cite{Ma2021}, yielding $\ell_2 = 1$ here.

\subsection{Re-scaling of the far field image}

\begin{figure}
    \centering
    \includegraphics[width=8cm]{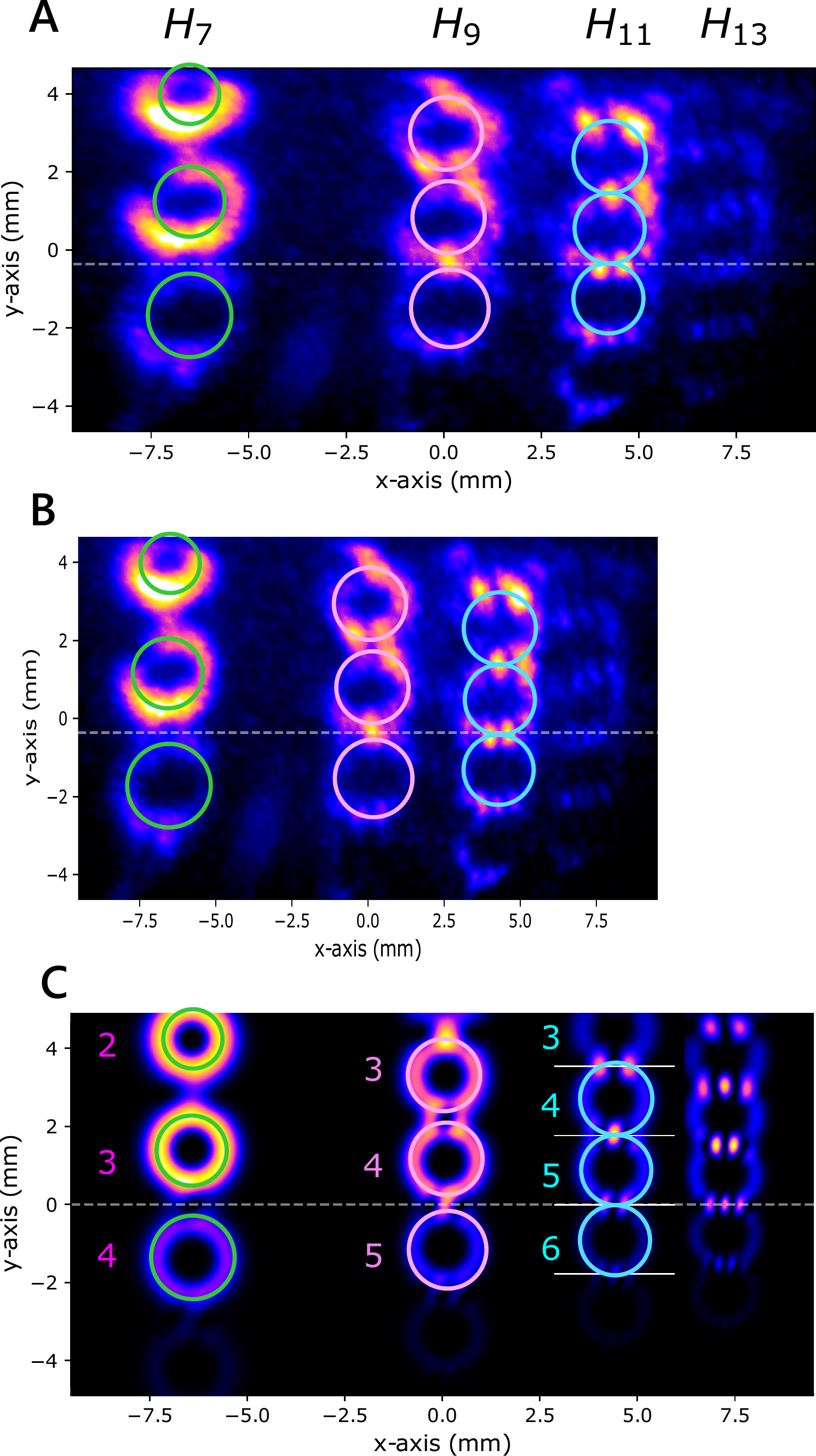}
    \caption{Rescaling of the far field images. \textbf{A} Raw image. \textbf{B} Image shrinked horizontally by a factor 0.8. \textbf{C} Comparison with simulation. The nine circles highlight the good agreement between the experiment and the simulation. They have the same radii in all figures. Circles of the same color are arranged equally in all figures. }
    \label{fig:rescaling}
\end{figure}

The micro-channel plates surface is not normal to the harmonics scattered by the grating. Thus, the far field image is slightly distorted along the horizontal direction. This can be seen clearly in Fig. SM\,\ref{fig:rescaling}.A, where the horizontal diameter of the optical vortices is larger than the vertical diameter. Such an effect is problematic if one tries to extract information about the spatial frequencies of the inter-vortices interference patterns. Since the unit of distance $\Delta$ (corresponding to half the distance between two neighboring vortices, for a given harmonic order) is measured along the vertical direction, it is mandatory that both dimensions of space have equal units. In order to compensate the distortion, we shrink the $x$ axis by a factor of 0.8, which corresponds to a tilt of 36 degrees. The result is shown in Fig. SM\,\ref{fig:rescaling}.B. The re-scaled harmonic vortices now have equal vertical and horizontal diameters. The ring diameters are very close to the ones expected from the simulation (Fig. SM\, \ref{fig:rescaling}.C). The Fourier transform of the line-outs of the interference fringes is then computed with the rescaled $x$ axis.

\subsection{Interference of two linearly polarized XUV vortices}
\label{sec: supp mat 1}

\begin{figure}
    \centering
    \includegraphics[width=4cm]{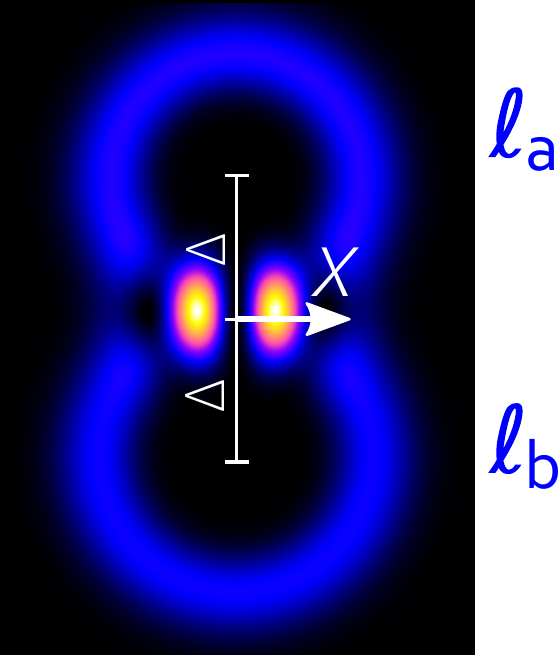}
    \caption{Simulated intensity pattern of two interfering XUV vortices,  with $\ell_a=3$ and $\ell_b=4$. Their relative propagation angle is adjusted to get the upper part of the lower one to interfere with the lower part of the upper one. }
    \label{fig:OAM interf}
\end{figure}

Let us consider two XUV vortex beams of identical frequencies and amplitudes $E_0$, aligned vertically (see Fig. SM\,\ref{fig:OAM interf}). Their centers are separated by a distance $2\Delta$. They respectively carry the OAM $\ell_a$ and $\ell_b$. We denote $x$ the coordinate along the horizontal line equidistant from the two vortex centers (white arrow in Fig. SM\,\ref{fig:OAM interf}). For small values of $x$ (i.e. $x \ll \Delta)$, the field along this line writes

\begin{equation}
\label{eq: fringe pattern}
    (E_1 + E_2)(x) \approx E_0 \big(e^{i \ell_a \frac{x}{\Delta}} + e^{-i \ell_b \frac{x}{\Delta} + i\phi}\big),
\end{equation}
with $\phi$ a relative phase which depends on the relative phase of the two vortices under consideration. The XUV intensity thus varies along the horizontal line as 

\begin{equation}
\label{eq: I fringe pattern}
    I(x) \approx E_0^2 \bigg(2+ 2\cos \bigg(\frac{x}{\Delta}(\ell_a+\ell_b) - \phi \bigg) \bigg).
\end{equation}
From this equation, we deduce the spatial frequency peak of the interference pattern: 

\begin{equation}
\label{eq: spatial freq}
    f_x \approx \frac{\ell_a+\ell_b}{2 \pi \Delta}.
\end{equation}
In our experiment, the two interfering vortices have consecutive OAM values, thus we consider the case $\ell_a = \ell$, $\ell_b = \ell +1$ and we obtain

\begin{equation}
\label{eq: spatial freq final}
    f_x \approx \frac{2\ell+1}{2 \pi \Delta}.
\end{equation}
The above formula yields $f_x(\ell=3)\cdot \Delta = 1.11$, $f_x(\ell=4)\cdot \Delta = 1.43$, $f_x(\ell=5)\cdot \Delta = 1.75$, etc. Comparing with Fig.\,2 of the main text, we observe that Eq.\,\ref{eq: spatial freq final} slightly overestimates the position of the spatial frequency components (found to be $f_x(\ell=3)\cdot \Delta = 0.93$, $f_x(\ell=4)\cdot \Delta = 1.28$, $f_x(\ell=5)\cdot \Delta = 1.51$)  although giving the correct trend. This is due to the fact that the formula derived here is only valid for small values of $x$. In practice, the Fourier transform is computed over a $2 \Delta$ wide line-out, which tends to decrease the spatial frequencies.

It is worth pointing out that in this specific experiment, we are observing the interference of two HHG photon channels corresponding to the same harmonic, which allows a direct retrieval of their relative phase $\phi$ with attosecond precision, since the temporal period of harmonic 11 is about 240 as.

\subsection{Line-outs of the vortex interference for harmonic 9 and 13}

\begin{figure}
    \centering
    \includegraphics[width=11cm]{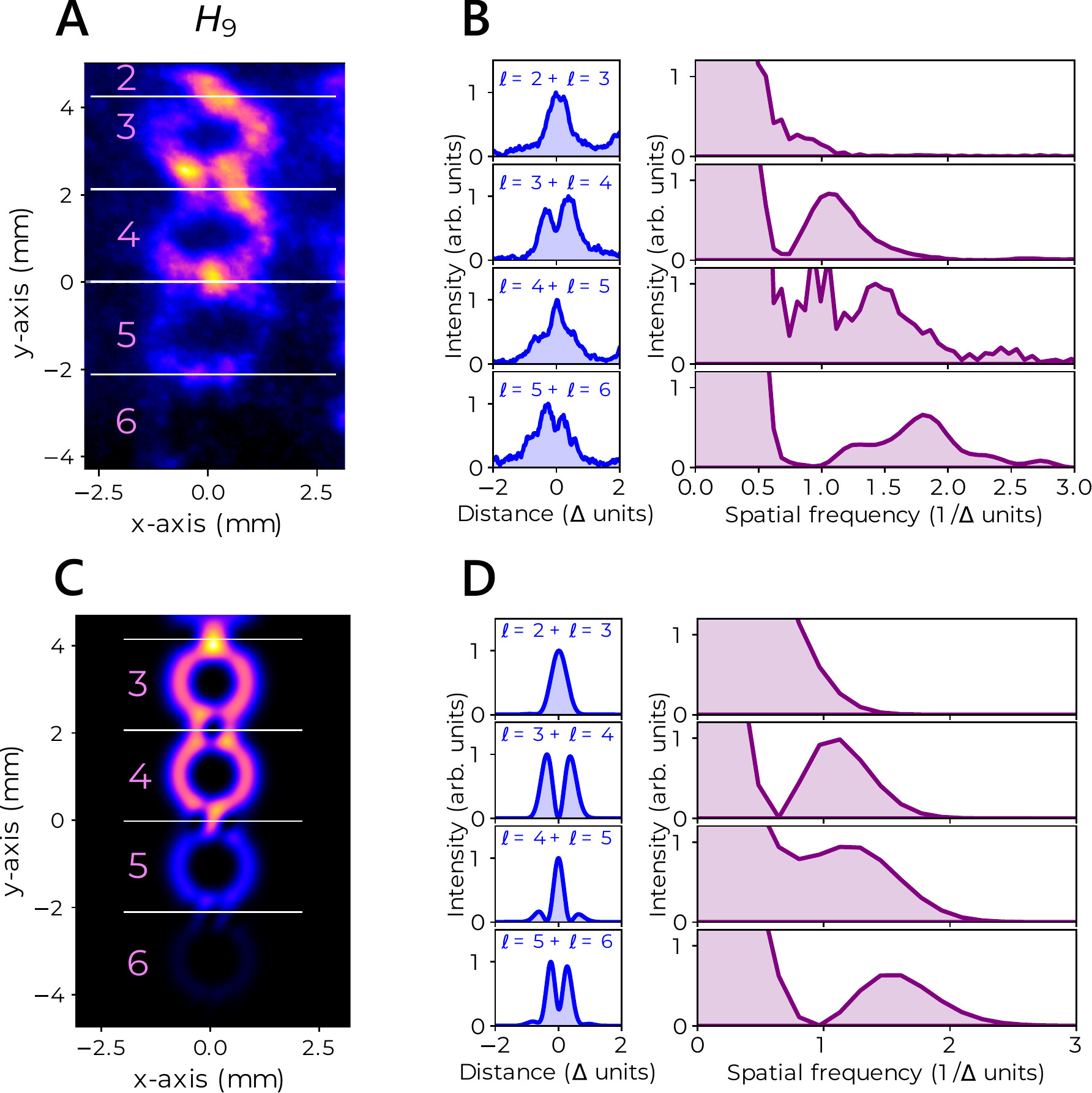}
    \caption{Analysis of the profile of harmonic $9$ generated by a combination of $\ell_1=0$ and $\ell_2 = 1$ linearly polarized non-collinear driving beams. \textbf{A} Experimental intensity profiles of harmonic $9$.  The colored digits indicate the OAM charge of the nearest XUV vortex.
    \textbf{B} Line-outs of harmonic 9 along the white horizontal lines in A (left), and corresponding spatial Fourier transform (right). 
   \textbf{C} Simulated intensity profiles of harmonic $9$.  \textbf{D} Line-outs of harmonic 9 along the white horizontal lines in C (left), and corresponding spatial Fourier transform (right).}
    \label{fig:Lineouts H9}
\end{figure}

\begin{figure}
    \centering
    \includegraphics[width=11cm]{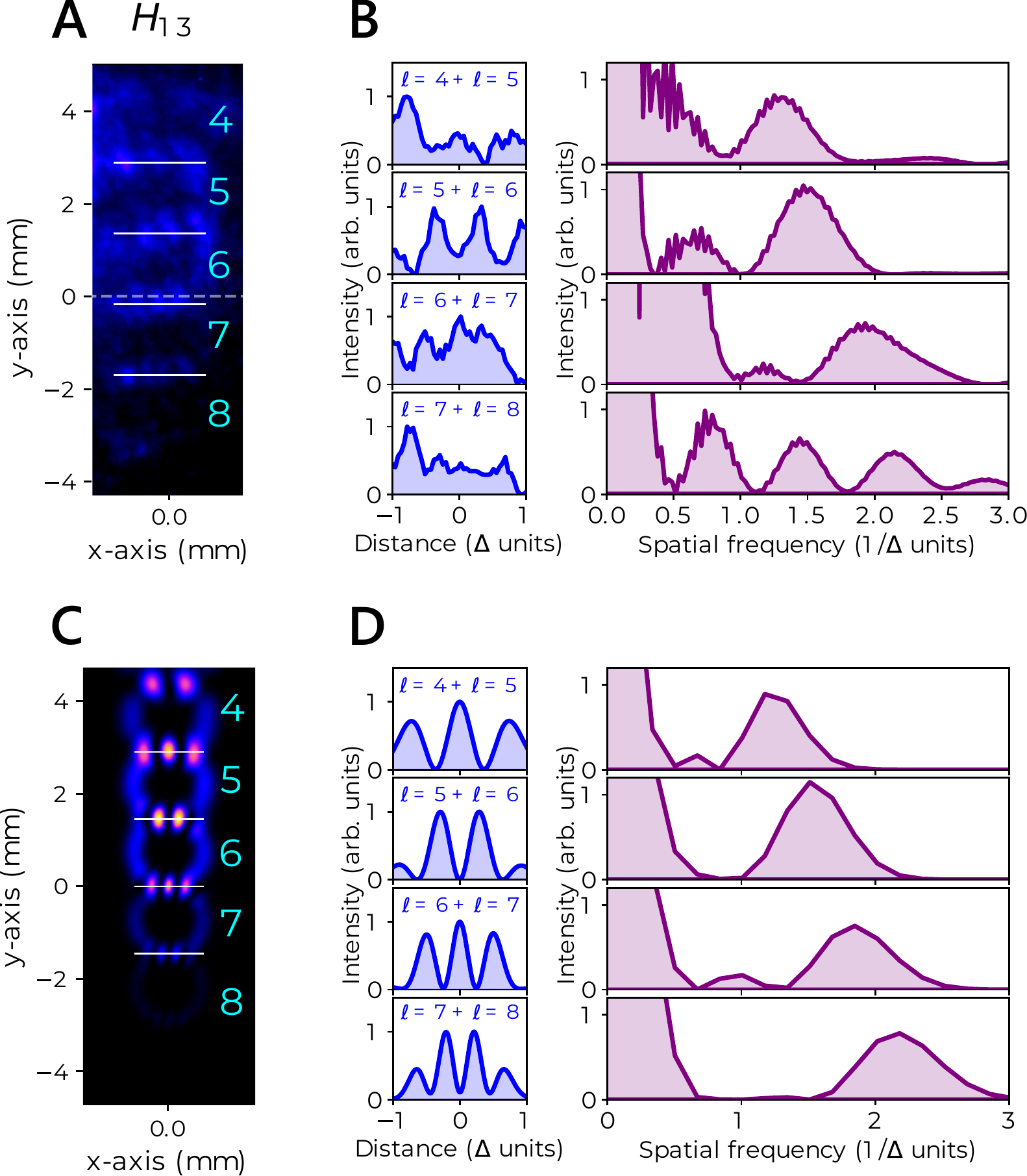}
    \caption{Analysis of the profile of harmonic $13$ generated by a combination of $\ell_1=0$ and $\ell_2 = 1$ linearly polarized non-collinear driving beams. \textbf{A} Experimental intensity profile of harmonic $13$.  The colored digits indicate the OAM charge of the nearest XUV vortex.
    \textbf{B} Line-outs of harmonic 13 along the white horizontal lines in A (left), and corresponding spatial Fourier transform (right). Due to the low signal to noise ratio of harmonic 13, the intensity patterns are multiplied by a half-period sine envelope before applying the Fourier transform. 
   \textbf{C} Simulated intensity profiles of harmonic $13$.  \textbf{D} Line-outs of harmonic 13 along the white horizontal lines in C (left), and corresponding spatial Fourier transform (right).}
    \label{fig:Lineouts H13}
\end{figure}

In the main text, the inter-vortex interference patterns of harmonic 11 are analyzed. The patterns are also visible for harmonic 9 and 13, even though the vortices overlap is not optimized for those harmonics. The analysis of harmonic 9 and 13 is displayed in Figs. SM\,\ref{fig:Lineouts H9}-\ref{fig:Lineouts H13}. The experimental and simulated line-outs, and their Fourier transforms, again show good agreement.

\subsection{Gaussian counter-rotating circularly polarized drivers}

\begin{figure}
    \centering
    \includegraphics[width=8cm]{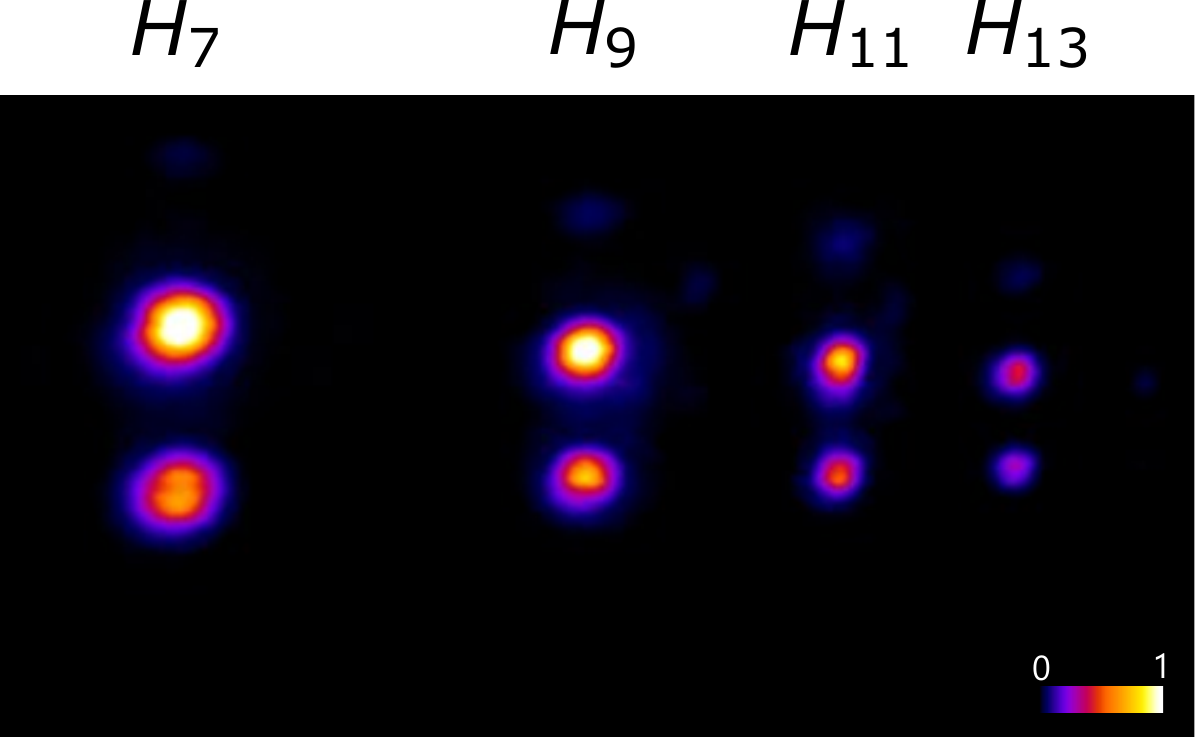}
    \caption{Intensity profiles of high harmonics generated by a combination of  counter-rotating circularly polarized non-collinear gaussian IR beams}
    \label{fig:hickstein}
\end{figure}

As a test of our setup, we generated harmonics with two non-collinear gaussian driving beams. The harmonic emission obeys the conservation of the photon spin, i.e. only two beamlets are observed for each harmonic (Fig. SM\,\ref{fig:hickstein}). We reproduce the results obtained by Hickstein \textit{et al.} \cite{Hickstein2015}. One can notice a weak third beamlet above the two main ones, which is due to the fact that the polarization of our driving beams is not perfectly circular. The far field profile is very sensitive to the waveplate angle: rotating the QWP or the HWP (see Fig. SM\,\ref{fig:setup}) by one degree is enough to make additional beamlets appear. We optimize the waveplate angle by minimizing the intensity of these additional beamlets.


\subsection{Partial LG to HG mode conversion}
\label{sec: supp mat 2}

\begin{figure}
    \centering
    \includegraphics[width=8cm]{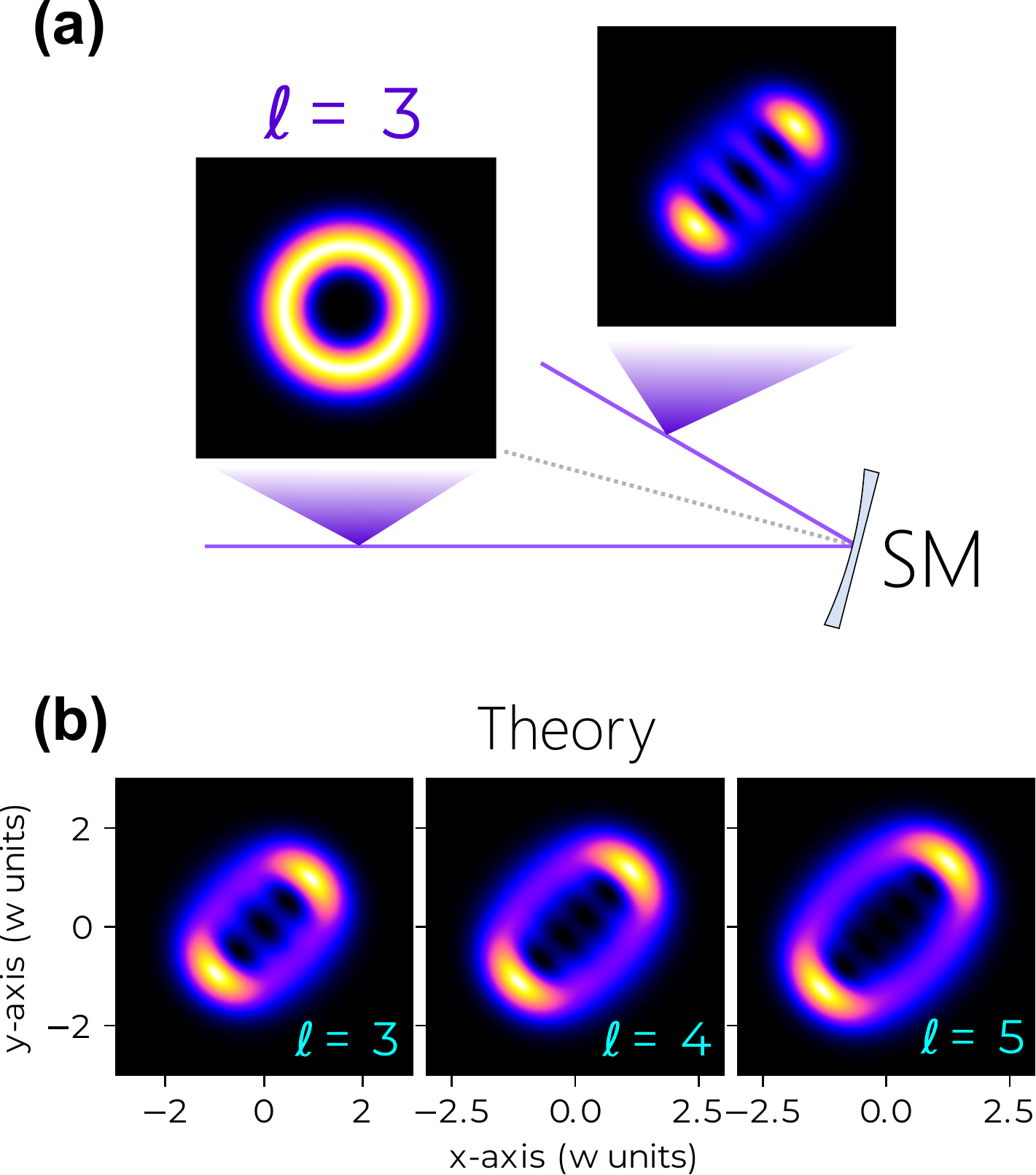}
    \caption{\textbf{a} Principle of the measurement of the OAM charge by partial LG to HG mode conversion. An incident $\ell=3$ LG mode with null radial index is partially converted in a $(n=3, m=0)$ HG mode tilted by 45\textdegree. \textbf{b} Numerically computed output mode of a $\Theta = 0.5$ rad mode converter, for an input LG mode with various azimuthal index $\ell$ (indicated in the bottom right corner), and a null radial index. $w$ is the beam width. The number of intensity nodes on the diagonal of each intensity profile corresponds to the absolute value of the OAM charge of the incident beam, and the inclination angle to its sign.}
    \label{fig:SMHGLGconverter}
\end{figure}

We chose the position of the gas jet to obtain XUV vortices very similar to single pure LG modes \cite{Geneaux2017}. The latter can be expressed as a sum of HG modes:

\begin{equation}
\text{LG}_{\ell, \rho} = \sum\limits_{n=0}^{+\infty} \sum\limits_{m=0}^{+\infty} a_{n, m}^{\ell, \rho}\text{HG}_{n,m}. 
\end{equation}
The expression of the $a_{n, m}^{\ell, \rho}$ coefficients can be found in \cite{Kimel1993}, (see Eq.\,4 and Eq.\,5). They are null for $m+n \neq 2\rho + |\ell|$.
HG modes with indices $n,m$ transmitted though an astigmatic optical system experience a dephasing due to the Gouy phase shift that is equal to a parameter $\Theta$ - determined by the experiment - times $m$. This process partially converts an incoming LG mode of OAM $\ell$ and null radial index into the corresponding HG $(n=\ell, m=0)$ mode, tilted by 45\textdegree. Such field is called an Ince-Gauss mode \cite{Bandres2004}. Ince-Gauss modes correspond to continuous transitions between the HG and LG modes, the same way elliptical polarization is a transition between linear and circular polarization.
The output field of the $\Theta$ converter writes
\begin{equation}
\text{C}_{\ell, \rho}(\Theta) = \sum_{n=0}^{+\infty} \sum_{m=0}^{+\infty} a_{n, m}^{\ell, \rho} e^{im\Theta}\text{HG}_{n,m}    
\end{equation}
Fig. SM\,\ref{fig:SMHGLGconverter}.b shows the output intensity of a partial mode converter with $\Theta = 0.5$ rad, for incident LG modes with various OAM charges and a null radial index. The intensity profiles are very close to the experimentally observed ones (see Fig.\,3 of the main text). The number of intensity minima on the diagonal of the image corresponds to the absolute value of the OAM of the incident vortex, while the orientation of the Ince-Gauss mode indicates its sign. 

\subsection{Mode conversion with linearly polarized harmonics}

\begin{figure}
    \centering
    \includegraphics[width=16cm]{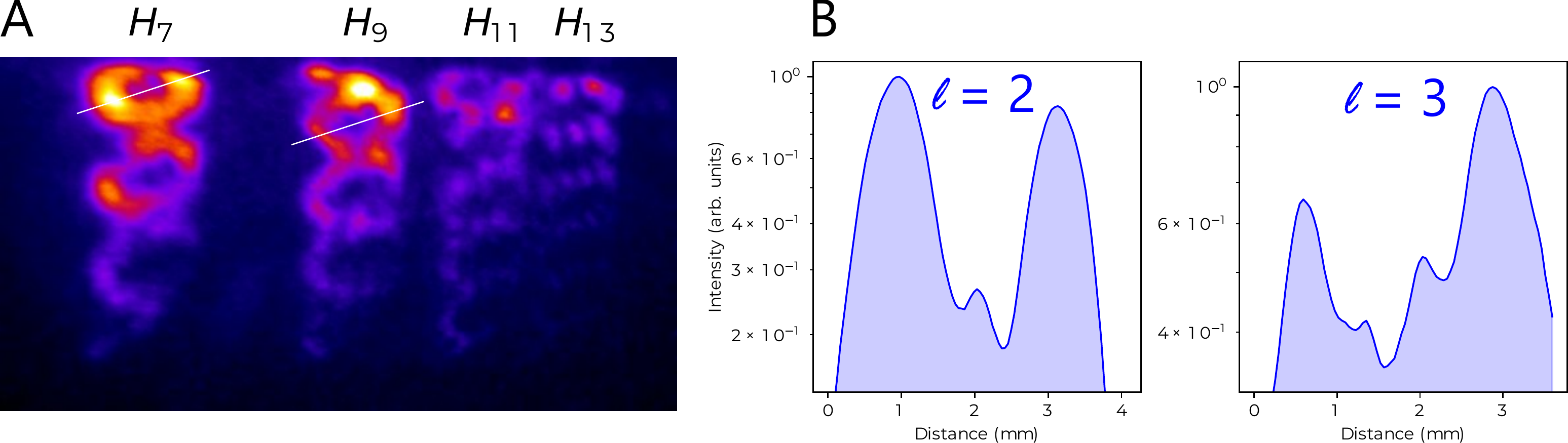}
    \caption{Partial LG to HG mode conversion of high harmonics generated from the non-collinear combination of two linearly polarized infrared drivers with OAM $\ell_1 = 0$ and $\ell_2=1$. \textbf{A} Far field intensity. \textbf{B} Line-outs along the white lines in A}
    \label{fig:astigmatic linear}
\end{figure}

Fig.\,3 of the main text shows the result of the LG to HG mode conversion, for circularly polarized harmonic beamlets. Fig. SM\,\ref{fig:astigmatic linear} shows the far field image obtained in the same conditions, but with vertically polarized driving beams. In this case, the mode conversion is also visible for the beamlet (5,2) of harmonic 7 (expected to carry an topological charge of 2, since it is emitted after the absorption of 5 photons from beam 1 and 2 photons from beam 2) and for the beamlet (7,3) of harmonic 9 (expected to carry a topological charge of 3). Indeed, respectively 2 and 3 intensity minima are observed on the diagonal of these beamlets.

\subsection{Spatio-temporal profile of the HHG emission in the non-collinear scheme}

\begin{figure}
    \centering
    \includegraphics[width=14cm]{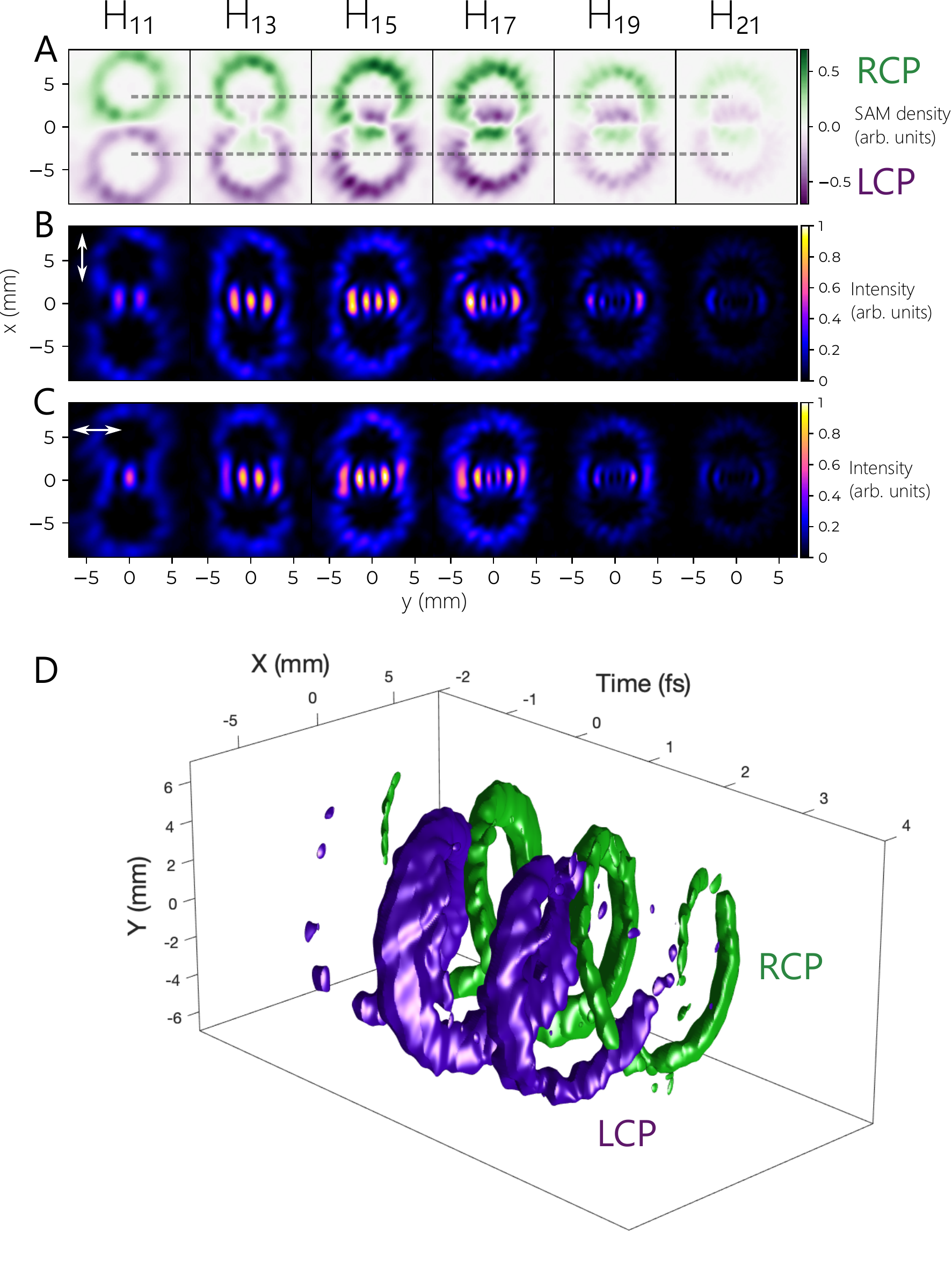}
    \caption{Results of SFA simulation of HHG with a polarization Möbius strip in non-collinear geometry. \textbf{A} SAM density for harmonics 11 to 21. Green and purple correspond to right and left circularly polarized light. Horizontal dashed lines indicate the good spatial overlap of the high harmonic vortices of same helicity. \textbf{B} Intensity profile after a vertical polarizer. \textbf{C} Intensity profile after a horizontal polarizer. \textbf{D}  Spatio-temporal intensity of the attosecond emission. Green and purple isosurfaces correspond to right and left circularly polarized attosecond pulses. The isosurface value is half of the maximum intensity}
    \label{fig:SFA}
\end{figure}

In order to obtain the spatio-temporal structure of the XUV emission, we perform simulations based on the strong field approximation (SFA), without resorting to the saddle-point approximation and including macroscopic phase-matching through the electric field propagator \cite{Hernandez-Garcia2010}. For practical reasons related to the computation time, the parameters of the simulation do not match the experimental ones. We consider for the simulation two equally intense non-collinear driving beams modelled as Laguerre-Gauss modes of topological charges 0 and 1. The total non-collinear angle is chosen to be 100 mrad (in order to obtain far field images similar to the experimental ones) and the beam waist is 30 µm. 
Each laser pulse is modelled with a $\sin^2$ envelope of 7.7 fs FWHM in intensity and 0.8 µm central wavelength. The total peak intensity is $2\times 10^{14}$ W/cm$^2$. The results are shown in Fig.SM\,\ref{fig:SFA}.

The $q\textsuperscript{th}$ harmonic is composed of two spatially separated beamlets with opposite $\pm 1$ helicity, carrying the OAM charge $(q\mp1)/2$ (Fig.SM\,\ref{fig:SFA}.A). In the region where the two beamlets overlap, the polarization direction rotates with respect to the horizontal coordinate. This periodic modulation becomes visible when plotting only the linear vertical or horizontal components of the XUV field (Fig.SM\,\ref{fig:SFA}.B,C), in the form of out of phase interference fringes.

Harmonic beamlets of same helicity show a rather good spatial overlap, as they are all emitted close to the bisector axis between the two driving beams. It is known that a superposition of different harmonics carrying an increasing OAM charge yields an attosecond light spring in the time domain \cite{Hernandez-Garcia2013, Geneaux2016}. The attosecond structure of our XUV field is shown in Fig.SM\,\ref{fig:SFA}.D. Indeed, two spatially separated light springs are observed, with right (green) and left (purple) circular polarization respectively. Compared to the light springs previously observed with HHG driven by linearly polarized LG beams \cite{Hernandez-Garcia2013, Geneaux2016}, which consisted in a set of two intertwined helices, each of our attosecond light spring, in addition of  presenting circular polarization, shows a unique helix. This specificity is due to the fact that the OAM charge scales as $(q-1)/2$ for the right circularly polarized vortex, and as $(q+1)/2$ for the left circularly polarized vortex, thus the derivative of the OAM charge with respect to the harmonic order is $1/2$ \cite{Turpin2017}, compared to $1$ in the case of HHG driven by a $\ell=1$ linearly polarized LG beam. 

This result can also be retrieved by observing the IR electric field at a given instant of time (Fig.\,1.A,B of the main text). There is only one azimuth where the field amplitude is maximum, i.e. the field is aligned with the major axis of the polarization ellipse. As time flows, the position of the peak amplitude rotates, achieving one full turn over an IR optical cycle. Thus, attosecond pulses are emitted in the form of a single-coil light spring. By contrast, the transverse profile of a linearly polarized LG driver of OAM $\ell=1$ exhibits two regions of peak amplitude, corresponding to a maximum and a minimum of the field's oscillation, resulting in a double-helix attosecond light spring.  

\bibliography{apssamp}
\newpage

